\documentclass{webofc}
\usepackage[varg]{txfonts}   
\usepackage{tikz}
\usepackage{svg}

\newcommand{\fatg}{{\rm{I}}\!\Gamma}
\begin{document}
\title{Three-gluon vertex in Landau-gauge from quenched-lattice QCD in general kinematics}
%
%

\author{ 
\firstname{F.} \lastname{Pinto-Gómez}\inst{1}\fnsep\thanks{\email{fpingom@alu.upo.es}} 
\and
\firstname{F.} \lastname{De Soto}\inst{1}\fnsep\thanks{\email{fcsotbor@upo.es}}
}

\institute{
Dpto. Sistemas F\'isicos, Qu\'imicos y Naturales, Univ. Pablo de Olavide, 41013 Sevilla, Spain
}

\abstract{%
We report on a novel and extensive lattice QCD analysis for the three-gluon vertex from quenched lattice-QCD simulations. Using standard Wilson action, we have computed the three-gluon vertex beyond the usual kinematic restriction to the symmetric $(q^2 = r^2 = p^2)$ and soft-gluon $(p = 0)$ cases where it depends on a single momentum scale. The so-dubbed bisectoral case $(r^2 = q^2 \ne p^2)$, where the transversely projected vertex can be cast in terms of three independent tensors, have been the object of a recent exhaustive scrutiny\,\cite{Pinto-Gomez:2022brg}, also shown in this communication. Herein, beyond this special case, results for kinematic configurations with three different squared momenta are also presented. All data considered, the lattice estimate of the three-gluon vertex exhibits a clear dominance of the tree-level tensor form factor. 
}
\maketitle
\section{Introduction}
\label{intro}

The intricate features of the three-gluon vertex, intrinsically related to the non-perturbative nature of the infrared dynamics of Quantum Chromodynamics (QCD), have been widely studied in the last few years\,\cite{Cucchieri:2006tf,Cucchieri:2008qm,Huber:2012zj,Pelaez:2013cpa,Aguilar:2013vaa,Blum:2014gna,Eichmann:2014xya,Mitter:2014wpa,Williams:2015cvx,Blum:2015lsa,Cyrol:2016tym,Athenodorou:2016oyh,Duarte:2016ieu,Boucaud:2017obn,Aguilar:2019uob,Aguilar:2021lke,Aguilar:2021okw,Catumba:2021yly,Catumba:2021hng,Sternbeck:2017ntv,Corell:2018yil,Aguilar:2019jsj,Aguilar:2019kxz,Vujinovic:2018nqc,Barrios:2022hzr,Pinto-Gomez:2022brg} and, importantly, connected with the emergence of a mass in the gauge sector the theory\,\cite{Aguilar:2008xm,Boucaud:2008ky,Fischer:2008uz,Dudal:2008sp,Tissier:2010ts,Cloet:2013jya,Pelaez:2014mxa,Eichmann:2021zuv,Gao:2017uox,Roberts:2021xnz,Binosi:2022djx,Roberts:2020udq,Roberts:2021nhw,Papavassiliou:2022wrb,Roberts:2020hiw}. The three-gluon vertex is a distinctive ingredient of QCD, triggered by its non-abelian nature and responsible for the appearance of asymptotic freedom at high energies and for the unavoidable need of implementation of non-perturbative methods to investigate its low-energy regime. The vertex itself can be computed using non-perturbative techniques, and a fruitful coordinate effort has been thus far developed, both using lattice-QCD and continuum methods, to unveil its structure, properties and main implications. 

Knowledge of three-gluon vertex is of great phenomenological relevance as, for instance, it is a key ingredient for continuum methods such as Dyson-Schwinger Equations (SDE) for gluon or quark propagator or quark-gluon vertex, or for Bethe-Salpeter Equations (BSE) for mesons or glueballs~\cite{Souza:2019ylx,Huber:2021yfy}.

In the present work, the transversely projected three-gluon vertex is computed using large statistics quenched SU(3) field configurations in the Landau gauge and for general kinematics. Previous studies have focused on either the case of one vanishing momentum (soft-gluon limit) or three equal squared momenta (symmetric limit). More recently, we have furthermore presented results\,\cite{Pinto-Gomez:2022brg} for kinematic configurations belonging to the denominated bisectoral case, in which two squared momenta are the same but differ from the third one ($r^2 = q^2 \ne p^2$). We report these results on this note, but also extend them by considering configurations with three different squared momenta, covering thereby the full kinematic domain available to the three-gluon vertex.

\section{Kinematics of the three-gluon vertex}
\label{sec-1}

The three-point Green's function is the correlation function  $\langle \widetilde{A}_\alpha^a (q)\widetilde{A}_\mu^b (r) \widetilde{A}_\nu^c (p)  \rangle$, with $\widetilde{A}_\alpha^a (q)$ standing for a SU(3) gauge field in Fourier-space with color index $a$, Lorentz index $\alpha$ and momentum $q$.
Focusing on the color asymmetric part of the Green's function, we will compute the quantity:
\begin{equation}\label{eq:Green}
    \mathcal{G}_{\alpha\mu\nu}(q,r,p) = \frac{1}{24} f^{a b c} \langle \widetilde{A}_\alpha^a (q)\widetilde{A}_\mu^b (q) \widetilde{A}_\nu^c (q)  \rangle
\end{equation}
from which the transversely projected vertex, $\overline{\Gamma}_{\alpha\mu\nu}(q,r,p)$, can be extracted as
\begin{equation}\label{eq:gamma}
    \mathcal{G}_{\alpha\mu\nu}(q,r,p) = g \overline{\Gamma}_{\alpha\mu\nu}(q,r,p)
    \Delta(q^2) \Delta(r^2) \Delta(p^2)
\end{equation}
where $\Delta(q^2)$ stands for the gluon propagator at a scale $q^2$ and $g$ is the SU(3) coupling constant.

\begin{figure}[t]
\begin{center}
\begin{tabular}{ccc}
\begin{tabular}{c}
\includegraphics[scale=0.3]{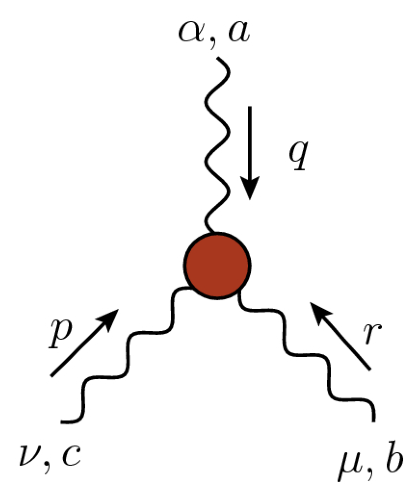} 
		\\
		\rule[0cm]{0cm}{3.5cm}	
\end{tabular}			
&		
\hspace*{-1.0cm}
\begin{tabular}{c}
\rule[0cm]{0cm}{1cm} \\
		\begin{tikzpicture}[scale=0.8]
			\node (r) at ( 3.0,  0.0) {}; %
			\node (q) at (-3.0,  0.0) {}; %
			\node (p) at ( 0.0, 5.36) {}; %
			\node (p2) at ( 0.0, 3.55) {}; 
			\node (r2) at (1.52, 0.87) {};
			\node (q2) at (-1.52, 0.87) {};					
			\node (S) at (0,1.75) {};
			\node (Op) at (0,2.625) {};
			\node (Oq) at (-0.76,1.31) {};
			\node (Or) at (0.76,1.31) {};
			\node (SGp) at (0,0) {};
			\node (SGq) at (1.52,2.63) {};
			\node (SGr) at (-1.52,2.63) {};
			\node (apA) at ( 3.0,  1.75) {}; %
			\node (amA) at ( -3.0,  1.75) {}; %
			\node (apB) at ( 0,  6.25) {};
			\fill[fill=blue!20] (p.center) -- (r.center) -- (q.center) -- (p.center);	-
			\draw [thick, blue] (r) -- (q) -- (p) -- (r);
			\fill[fill=white] (S) circle (1.72cm);
			\draw [very thick,gray] (p2) -- (SGp);
			\draw [thick,black,dashed] (r2) -- (SGr);
			\draw [thick,black,dashed] (q2) -- (SGq);
			\node at (-0.25,2.075) {$S$};
			\node at (S)[circle,fill,inner sep=1.5pt,green]{};
			\node at (0,-0.4) {${p^2=0}$};
			\node at (SGp) [circle,fill,inner sep=1.5pt,orange]{};
			\node at (2.5,2.8) {${q^2=0}$};
			\node at (SGq) [circle,fill,inner sep=1.5pt,black]{};
			\node at (-2.5,2.8) {${r^2=0}$};
			\node at (SGr) [circle,fill,inner sep=1.5pt,black]{};
		\end{tikzpicture} 
\end{tabular}
&
\hspace*{-1.0cm}
\begin{tabular}{c}
		\begin{tikzpicture}[scale=0.4]
			\fill[fill=blue!30] (0,3) -- (3,0) -- (-1.25,-1.25) -- (0,3);
			\draw [very thick, -latex] (0,0) -- (0,5) node [right] {$p^2$};
			\draw [very thick, -latex] (0,0) -- (-2.5,-2.5) node [right] {$q^2$};
			\draw [very thick, -latex] (0,0) -- (5,0) node [right] {$r^2$};
			\draw [thick, blue] (0,3) -- (3,0) -- (-1.25,-1.25) -- (0,3);
			\draw [thick, blue, -latex] (0,0) -- (3.5,3.5);
			\node at (0.75,0.75) [circle,fill,inner sep=1.2pt,green]{};
		\end{tikzpicture}
		\\
		\rule[0cm]{0cm}{3.5cm}		
\end{tabular}		 
\end{tabular}
\end{center}
\caption{The kinematic configuration of the three-gluon vertex (left diagram) represented by the Cartesian coordinates $(q^2,r^2,p^2)$ (right picture). All kinematics allowed are contained in a circle around the symmetric case ($q^2=r^2=p^2$, green dot). The bisectoral line for $r^2=q^2$ (thick gray), and the particular soft-gluon (orange solid circle), and symmetric (green) cases appear depicted. The other two  bisectoral lines (black dashed) and their corresponding soft-gluon limits (black dots) are also illustrated.  
}
\label{fig:triangle}
\vspace*{-0.5cm}
\end{figure}
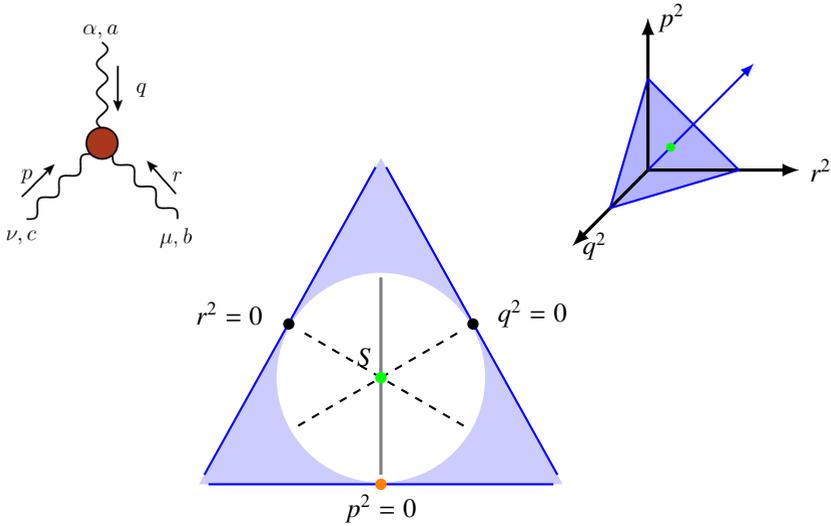

The three-gluon vertex depends on the three incoming momenta, $q$, $r$ and $p$, which, owing to momentum conservation, are restricted by  $q+r+p=0$. Therefore, any scalar form factor can be written as a function of $q^2$, $r^2$ and $p^2$; and each kinematic configuration can be represented as a point in the positive octant of a three-dimensional space with axes $q^2$, $r^2$ and $p^2$, with all symmetric configurations ($q^2=r^2=p^2$) occupying the octant diagonal. 
Alternatively, the vertex form factors can be also written in terms of angles. For example, one may choose two squared momenta (let them be $q^2$ and $r^2$) and their angle, defined by $\cos{\theta_{qr}}=(p^2-q^2-r^2)/2\sqrt{q^2 r^2}$, instead of the third squared momenta, $p^2$. The following symmetric combination of momenta, 
\begin{equation}\label{eq:s2}
s^2=(q^2+r^2+p^2)/2\;,
\end{equation}
can be defined such that all the configurations sharing the same value of $s^2$ lie on a plane perpendicular to the octant diagonal, the distance from the plane to the origin given by $2s^2/\sqrt{3}$. Furthermore, a second combination of momenta, 
\begin{equation}\label{eq:t2}
    t^2 = \sqrt{(q^2-r^2)^2+(r^2-p^2)^2+(p^2-q^2)^2} / \sqrt{3} \;, 
\end{equation}
can be seen to measure the distance from the symmetric point to the one representing a given configuration, within the plane characterized by its $s^2$-value. Momentum conservation imposes an upper bound for this distance, $t^2 \leq \sqrt{2/3} s^2$ (white circle in Fig.~\ref{fig:triangle}), thus defining a cone around the octant diagonal containing all possible kinematic configurations for the three-gluon vertex. As an illustration, one can move along the line for the bisectoral case, $q^2=r^2$ (thick gray line in Fig.~\ref{fig:triangle}), from the soft-gluon limit, $p^2=0$ (orange dot), up to the co-linear limit reached at $p^2=4q^2$ (which corresponds to $q=r=-p/2$). In terms of the angle, varying $p^2$ from $0$ to $4q^2$ along the bisectoral line implies for the angle $\theta_{qr}$ its decreasing from $\pi$ to  $0$. 

As will be exposed below, one can capitalize on the Bose symmetry to characterize the three-gluon vertex through scalar form factors expressed in terms of $s^2$ and $t^2$, supplemented by an additional third symmetric combination of momenta\,\cite{Eichmann:2014xya,Pinto-Gomez:2022brg}.

\section{A Bose-symmetric tensor representation}

The one-particle irreducible (1PI) three-gluon vertex can be decomposed into a pole-free component and a term only comprising the massless longitudinal poles triggering the Schwinger mechanism for gluon mass generation, $\fatg_{\alpha\mu\nu}=\Gamma_{\alpha\mu\nu} + V_{\alpha\mu\nu}$\,\cite{Aguilar:2011xe,Ibanez:2012zk,Binosi:2017rwj}. The pole-free component, $\Gamma_{\alpha\mu\nu}$, can be expanded in a full basis consisting of $14$ tensors\,\cite{Ball:1980ax}:
\begin{equation}
    \Gamma_{\alpha\mu\nu}(q,r,p) = \sum_{i=1}^{10} X_i (q^2,r^2,p^2) \ell_i(q,r,p) + \sum_{i=1}^{4} Y_i (q^2,r^2,p^2) {t}_i(q,r,p)\,,
\end{equation}
four of them transverse, $t_i$, and the rest, $\ell_i$, non-transverse. 
Through a lattice QCD simulation, one can only access the transversely projected vertex, Eq.~(\ref{eq:gamma}), which is the transverse projection of the full 1PI vertex:
\begin{eqnarray}
     \overline{\Gamma}_{\alpha\mu\nu}(q,r,p) = {\fatg}_{\alpha\mu\nu}(q,r,p) P_{\alpha\alpha'}(q) P_{\mu\mu'}(r) P_{\nu\nu'}(p) 
   =  \Gamma_{\alpha\mu\nu}(q,r,p) P_{\alpha\alpha'}(q) P_{\mu\mu'}(r) P_{\nu\nu'}(p) \,, 
\end{eqnarray}
where $P_{\alpha\alpha'}(q)=g_{\alpha\alpha'}-q_\alpha q_{\alpha'}/q^2$. 
As $\overline{\Gamma}_{\alpha\mu\nu}(q,r,p)$ is transverse, it can be expressed only as a combination of the four transverse tensors $t_i$, although a more suitable representation follows instead from the use of a fully Bose-symmetric basis. The Bose symmetry implies for $\overline{\Gamma}_{\alpha\mu\nu}$, defined by  Eqs.\,(\ref{eq:Green},\ref{eq:gamma}) as the contraction of a Green's function with the fully antisymmetric tensor $f^{a b c}$, its reversing the sign under exchange of both gluon Lorentz indices and momenta, {\it e.g.} $(q,\alpha) \longleftrightarrow (r,\mu)$. We have then chosen the following basis tensors\,\cite{Pinto-Gomez:2022brg}:  
\begin{subequations}
\label{eq:tensors}
\begin{align}
\widetilde{\lambda}_1^{\alpha \mu \nu} \!=& \ P_{\alpha'}^\alpha(q)  P_{\mu'}^{\mu}(r)  P_{\nu'}^\nu(p) 
\left[\ell_1^{\alpha'\mu'\nu'} + \ell_4^{\alpha'\mu'\nu'} + \ell_7^{\alpha'\mu'\nu'}\right] \,,
\label{eq:tl1}
\\ 
\widetilde{\lambda}_2^{\alpha \mu \nu} \!=&  \frac{3}{2 s^2} \,(q-r)^{\nu'} (r-p)^{\alpha'} (p-q)^{\mu'} 
P_{\alpha'}^\alpha(q)  P_{\mu'}^{\mu}(r)  P_{\nu'}^\nu(p)\,,
\label{eq:tl2}
\\
\widetilde{\lambda}_3^{\alpha \mu \nu} \!=& \frac{3}{2 s^2}  P_{\alpha'}^\alpha(q)  P_{\mu'}^{\mu}(r)  P_{\nu'}^\nu(p) 
\left[\ell_3^{\alpha'\mu'\nu'} + \ell_6^{\alpha'\mu'\nu'} + \ell_9^{\alpha'\mu'\nu'}\right]\,,
\label{eq:tl3}
\\
\widetilde{\lambda}_4^{\alpha \mu \nu} \!=& \left( \frac{3}{2 s^2}\right)^2
\left[t_1^{\alpha\mu\nu} + t_2^{\alpha\mu\nu} + t_3^{\alpha\mu\nu}\right]\,;
\label{eq:tl4}
\end{align}
\end{subequations}
each obeying the same antisymmetric property as $\overline{\Gamma}_{\alpha\mu\nu}$,  
with $t_i$ and $\ell_i$ corresponding to the tensors of the so-called Ball-Chiu basis ({\it e.g.}, Eqs.\,(3.4,3.6) in \cite{Aguilar:2019jsj}). The transversely projected vertex can be then written as
\begin{equation}
    \overline{\Gamma}^{\alpha\mu\nu}(q,r,p) =\sum_{i=1}^4 \overline{\Gamma}_i(q^2,r^2,p^2) \widetilde{\lambda}_i^{\alpha \mu \nu} (q,r,p) \;,
\end{equation}
Bose symmetry guaranteeing that $\overline{\Gamma}_i(q^2,r^2,p^2)$ are symmetric under permutations of two momenta. In Eqs.~(\ref{eq:tl1}-\ref{eq:tl4}), $\widetilde{\lambda}_1^{\alpha \mu \nu}$ is chosen to be the tree-level tensor, which displays a non-zero component in all possible kinematic configurations. Given special configurations, the four tensors are not independent. Particularly, for soft-gluon cases, there is only one linearly independent tensor; for the symmetric case, there are two, usually taken as $\lim_{q^2=r^2=p^2} \widetilde{\lambda}_{1,2}^{\alpha \mu \nu}(q,r,p)$\,\cite{Boucaud:2018xup}; and for the bisectoral (represented by the vertical gray line in Fig.~\ref{fig:triangle}) there are three: the corresponding limit from the tensor \eqref{eq:tl4} can be easily recast as a linear combination of those from (\ref{eq:tl1}-\ref{eq:tl3})\,\cite{Pinto-Gomez:2022brg}. 

Bose symmetry enforces then the invariance of the form factors $\overline{\Gamma}_i(q^2,r^2,p^2)$ under exchange of momenta, such that they must be functions of a Bose-symmetric combination of $q^2$, $r^2$ and $p^2$, as $s^2$ or $t^2$ introduced earlier. A detailed derivation of the latter can be based on the properties of the irreducible representations of the permutation group\,\cite{Eichmann:2014xya}, or grounded on a geometrical analysis\,\cite{Pinto-Gomez:2022brg}.

\section{Results}

\begin{figure}[!h]
\centering
\includegraphics[width=\textwidth]{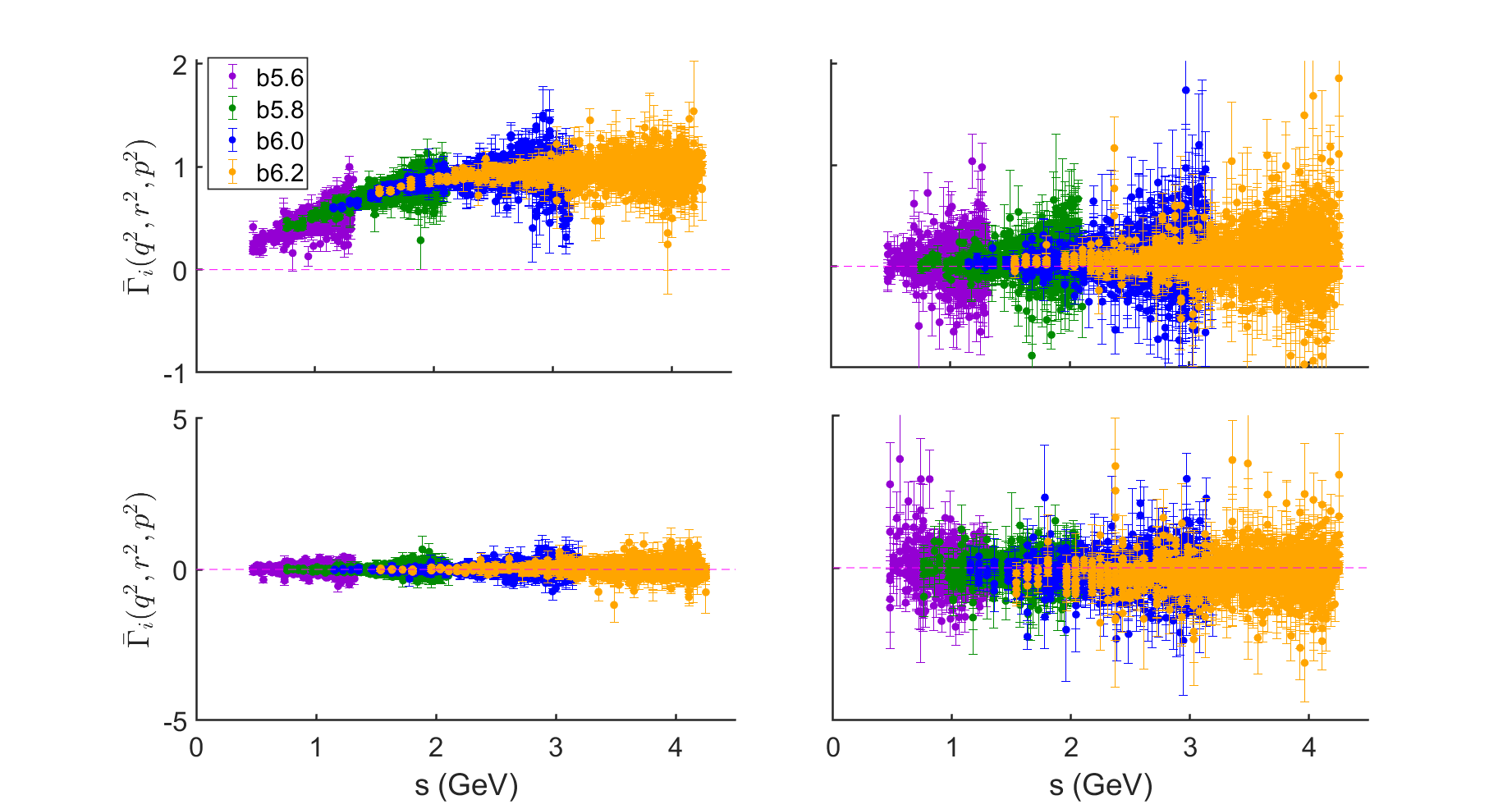} 

\caption{Form factors $\overline{\Gamma}_{1\, R}(q^2,r^2,p^2)$ to $\overline{\Gamma}_{ 4\, R}(q^2,r^2,p^2)$ obtained from the gauge-field configurations of Tab. \ref{configurations} plotted in terms of $s$, from top left to bottom right $\Gamma_1, \Gamma_2, \Gamma_3, \Gamma_4$. These plots show the kinematics that are not bisectoral, i.e., none of the momenta squared are equal. For these kinematics the four tensors (\ref{eq:tl1}-\ref{eq:tl4}) are independent. The kinematic configurations near $q^2=r^2$ (lying on the gray line in Fig. \ref{fig:triangle}) or the equivalent cases $r^2=p^2$ or $p^2=q^2$ or the kinematics where $t^4$ approaches its maximum value (lying near the circle in Fig.~\ref{fig:triangle} have been discarded. For those case, the projection matrix takes eigenvalues which approach zero, leading to very noisy results. The renormalization point is $\mu$=4.3 GeV.} 
\label{fig:fourtensorcase}
\end{figure}

The scalar form factors $\overline{\Gamma}_i(q^2,r^2,p^2)$ can be projected out from the lattice data  for $\overline{\Gamma}_{\alpha\mu\nu}(q,r,p)$ and renormalized by applying the standard \emph{momentum subtraction} scheme (MOM), as described in \cite{Pinto-Gomez:2022brg}. The corresponding form factors obtained for general kinematics have been represented in Fig.\ref{fig:fourtensorcase} as a function of $s^2$ and for a subraction point $\mu$=4.3 GeV. In the plot, about 3000 points have been displayed, representing each a kinematic configuration $(q^2,r^2,p^2)$ constrained by the following cuts: $|q^2-r^2|/s^2,|r^2-p^2|/s^2,|p^2-q^2|/s^2 < 0.35$, removing configurations near the three bisectoral lines in the triangle of Fig.\,\ref{fig:triangle}; and $t^2/s^2 < 0.7$, which prevents from taking the ones lying close to the circumference ($t^2=\sqrt{2/3} s^2$). The reason to exclude those kinematics is that their form factors have to be projected out by the inversion of a matrix taking eigenvalues approaching zero (in the bisectoral limit, for example, there are only three independent tensors instead of four), exhibiting a behavior compatible with that for selected kinematics but significantly increasing the statistical noise.

\begin{table}[h]
\centering
	\begin{tabular}{c c c c}
		\hline
		\hline
		$\beta$ & $L^4/a^4$ & a (fm) & confs \\
		\hline
		5.6 & $32^4$  & 0.236 & 2000 \\
		\hline
		5.8 & $32^4$  & 0.144 & 2000 \\
		\hline
		6.0 & $32^4$  & 0.096 & 2000 \\
		\hline
		6.2 & $32^4$  & 0.070 & 2000 \\
		\hline
		\hline
	\end{tabular}
	\caption{Quenched configurations employed for lattice calculations.}	
	\label{configurations}
\end{table}

Two outstanding results emerge from Fig.~\ref{fig:fourtensorcase}. 
the first is that only the form factor associated to $\widetilde{\lambda}_1$, the tree-level tensor, displays a clear non-zero signal. For the three others, one is left with a very poor signal to noise ratio, and they can be thus taken to deliver a null or negligible result. 
The second remarkable outcome is that the form factors are seen to depend only on the Bose-symmetric variable $s^2$:  when plotted in terms of this variable, the form factors for all the different combinations of $(q^2,r^2,p^2)$, whose geometric representation lie on a plane defined by the same $s^2$, take the same value. 
This property has been exposed and named thereby \emph{planar degeneracy} in Ref.\,\cite{Pinto-Gomez:2022brg}, through a systematic analysis of the three-gluon vertex from lattice QCD in the bisectoral kinematics (thick gray line of Fig.\,\ref{fig:triangle}). The latter is the first lattice study confirming this special behavior of the three-gluon vertex, which had been formerly introduced from a  continuum analysis with a particular truncation scheme\,\cite{Eichmann:2014xya}. 
The novelty of the results presented here is that they come from the first lattice study for general kinematics, exploring any class of configurations and the evaluation of the form factors for the four tensors, when they are independent. 

\begin{figure}[!h]
\begin{center}
\includegraphics[width=\textwidth]{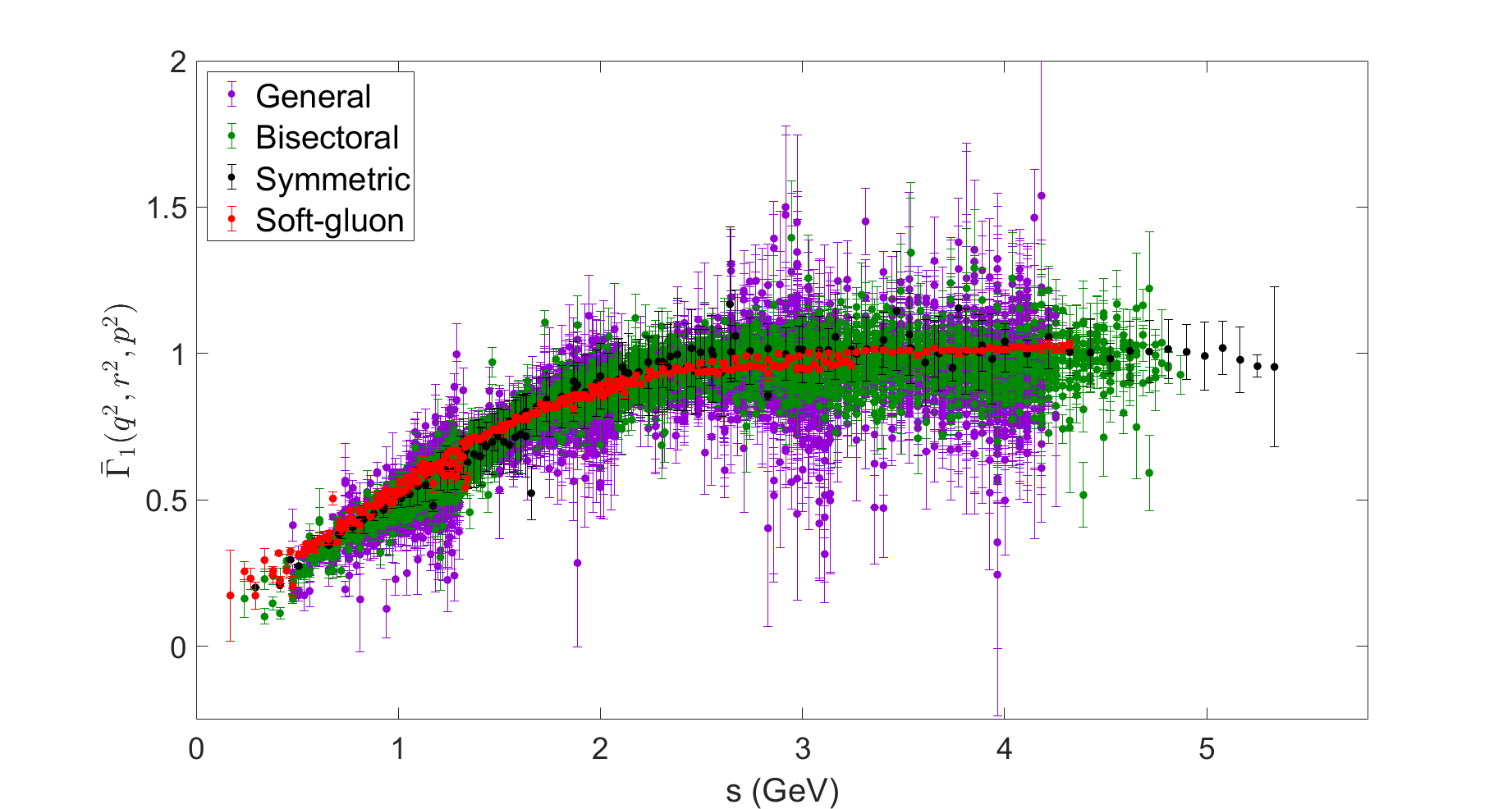} 
\end{center}
\caption{Illustration of the scaling of lattice data for the scalar form factor $\overline{\Gamma}_{1\, R}(q^2,r^2,p^2)$ in terms the symmetric combination of momenta $s$, for the different kinematics analyzed (in the legend). Each color represents the data for the four lattice setups in Tab.~\ref{configurations}.The normalization point is $\mu$ = 4.3 GeV.} 
\label{fig:gamma1}
\vspace*{-0.5cm}
\end{figure}

In the bisectoral case $q^2=r^2$, only three tensors become independent, the bisectoral limits of Eqs.\,(\ref{eq:tl1}-\ref{eq:tl3}), whose corresponding form factors have been evaluated in Ref.\,\cite{Pinto-Gomez:2022brg}. For the symmetric case, only contributions from $\widetilde{\lambda}_1$ and $\widetilde{\lambda}_2$ survive in Eqs.\,(\ref{eq:tl1}-\ref{eq:tl2}); while, in the soft-gluon cases ($p=0$, $r=0$ or $q=0$), only $\widetilde{\lambda}_1$ from Eq.\,\eqref{eq:tl1} contributes.  Therefore, as the form-factor $\overline{\Gamma}_1$ is the only present in all the special cases under consideration (soft-gluon, symmetric and bisectoral), being furthermore the one delivering a clear significant signal and a non-negligible result in general kinematics, we have chosen to display its lattice estimates for different kinematics, all together in Fig.\,\ref{fig:gamma1}. The plot makes remarkably apparent that, plotted in terms of $s^2$, the results for all the different kinematics compares very well, exhibiting the same behavior. The statistical spreading of the data for general kinematics becomes larger than those for the bisectoral case,  as can be expected for a number of accessible momenta triads increasing very fast when the kinematics is extended. A positive outcome of this fact would be, in principle, its providing a great set of data scaling in terms of $s^2$ for the application of H4-extrapolation methods\,\cite{Becirevic:1999uc,Becirevic:1999hj,deSoto:2007ht}, thus allowing for a more refined treatment of lattice discretization artifacts. All together, the feature highlighted by Figs.\,\ref{fig:fourtensorcase} and \ref{fig:gamma1} can be condensed in the following result:
\begin{equation}\label{eq:final}
\overline{\Gamma}^{\alpha \mu \nu}(q,r,p) \approx L_{sg}\left(\frac{q
^2+r^2+p^2}{2}\right) \widetilde{\lambda}_1^{\alpha\mu\nu}(q,r,p) \;,
\end{equation}
as a very suitable approximation for the transversely projected three-gluon vertex, where $\widetilde{\lambda}_1$ is given by Eq.\,\eqref{eq:tl1} and $L_{sg}$ stands for the only surviving form factor in the soft-gluon limit: $L_{sg}(s^2)\equiv \overline{\Gamma}_1(s^2,s^2,0)$.

\section{Conclusions}

We have presented a quenched lattice calculation of the transversely projected three-gluon vertex for general kinematics. Using a 
Bose-symmetric basis, we found that the tree-level tensor clearly dominates over the three others. Moreover, we have observed that the vertex form factors depend predominantly on the special variable $s^2=(q^2+r^2+p^2)/2$; i.e., the form factors take the same value for all kinematics configuration lying within the plane defined by $s^ 2=cte$ shown in Fig.~\ref{fig:triangle}. This {\it planar degeneracy} of the form factors, found empirically, has at the current stage no deeper understanding of its origin, but also appears in the context of DSE analysis~\cite{Eichmann:2014xya}, and its confirmation induces a remarkable simplification for physical applications that depend on the structure of the three-gluon vertex, namely Eq.\,\eqref{eq:final}. Particularly, it can be successfully applied to evaluate the so-called \emph{displacement function}\,\cite{Aguilar:2021uwa} from the lattice, a key ingredient showing the activation of the Schwinger mechanism for the gluon mass generation through the Slavnov-Taylor identity involving the three-gluon vertex.

\noindent\textbf{Acknowledgments}. 
We are grateful to M.N. Ferreira, J. Papavassiliou, J. Rodríguez-Quintero, and C.D. Roberts
for the helpful discussions. This
project was supported by Spanish research project PID2019-107844-GB-C2 and the regional Andalusian P18-FR-5057.
All calculations have been performed at the UPO computing center, C3UPO.

\end{document}